# Ordered Statistics Vertex Extraction and Tracing Algorithm (OSVETA)

Bata VASIC

*Faculty of Electronic Engineering, University of Nis, 18000 Nis, Serbia*
bata.vasic@elfak.ni.ac.rs

*Abstract*— We propose an algorithm for identifying vertices from three dimensional (3D) meshes that are most important for a geometric shape creation. Extracting such a set of vertices from a 3D mesh is important in applications such as digital watermarking, but also as a component of optimization and triangulation. In the first step, the Ordered Statistics Vertex Extraction and Tracing Algorithm (OSVETA) estimates precisely the local curvature, and most important topological features of mesh geometry. Using the vertex geometric importance ranking, the algorithm traces and extracts a vector of vertices, ordered by decreasing index of importance.

## I. Introduction

Curvature plays an important role in the global shape recognition, but there are others significant features of discrete object representations, which are involved in description of local topological characteristics. Thus, for example a low and high value of dihedral angles characterizes local topology ridge and furrow respectively. In the other hand, great distance of one point to its neighbors probably represents topological error, despite to evident high value of local curvature. Estimation of such features importance in combination with a local and global estimation of curvature is the main idea of this paper. In computers graphics 3D object is described in discrete form and fundamental element is a vertex given by its Euclidian coordinates. Therefore, the goal of our approach is a selection of important vertices, i.e. vertices that are most resistant to optimization and simplification processes.

For a best defining criteria it is necessary a theoretical study of existing curvature estimation algorithms for the beginning. Results will demonstrate the effect of the different curvature estimation to the description of topological features. Within next phase follows an exploration of most important characteristics of optimization and simplification processes. In order to decrease a number of vertices, the optimization process uses topological criteria that determine "irrelevant" vertices for deletion from the mesh. We need to explain those useful criteria as principles of the vertex removal process, thus we define and successfully eliminate such vertices from our selection. Actually, in the context of vertex assessment well understanding of simplification criteria determines criteria that should be give bad grades. Thus, our approach provides forming criteria for action from both sides of problem. First set of criteria eliminates vertices from ordering based on matching their geometrical features with criteria conditions. The purpose of second set of criteria is ordering and finally, determination of the vertex extraction order in our approach is in direct proportion to its invariance in relation to the simplification and optimization process.

The paper is organized as follows: Section II gives preliminaries with discussing a prior work in relevant areas of 3D geometry. In this section, we also define the notion of a curvature and its sign, and introduce the concepts and terminology used in the paper. Typical methods of optimization and study of their characteristic properties are also located in this section. Section III presents the main idea as well as the details of OSVETA extraction algorithm. This section also provides conclusions about requirements that assessment criteria have to meet. In Section IV we show numerical results, which are obtained using our OSVETA software tools with real 3D meshes. In this section, we also discuss the limitations and possible shortcomings of the proposed approach and tradeoffs among complexity, distortion and security. Section V provides the summary of the numerical results, conclusions and future research directions. In Appendix (A and B), one may find statistic results of experimental test carried out to determine the rating of certain criterion.

## II. Preliminaries

Estimation of curvature has long been unavoidable in the process of shapes perception and recognition [1], [2]. In 1954, Attneave [1] published an important paper in which he observed that extrema of curvature along contours provided much of the information necessary to recognize objects from line drawings. Attneave manually picked points that corresponded to extrema of curvature and connected the points with straight lines. Remarkably, recognition based on such figures was found to be simple for human observers. However, defining important vertices, which form a simple and computer-recognizable 3D shape, is not trivial process. Thus, it is necessary to first define the problems and difficulties of discrete curvature estimation, but also point out differences between known methods.

*A. Discrete curvature estimation methods*

The notion of curvature originates from Gauss and his work in differential geometry [3] where he was first who formulated the main curvature features. Further curvature evaluation [4] methods improve mathematical interpretation and introduce new conditions, generalizing computation for all types of 3D meshes [5] [6]. For example, boundary meshes, that Gauss did not consider in his aforementioned work. However, the main discussions are conducted about accuracy of methods for discrete mesh computations.

*1) Curvature Estimation using Differential Geometry*

From differential geometry [7] we know that for manifold surface $M$ in $\mathbb{R}^3$, and each point on the given surface, one can locally approximate the surface by its tangent plane that is orthogonal to the *normal vector* **n**. For every unit direction **e** in the tangent plane, the normal curvature **K** is defined as the curvature of the curve that belongs to both the surface itself and the plane containing both **n** and **e**. The two *principal curvatures:* $\kappa_1$ and $\kappa_2$ of the surface $M$, with their associated orthogonal directions $\mathbf{e}_1$ and $\mathbf{e}_2$ are extreme values of all normal curvatures (Fig. 1. a). The *mean curvature* $\kappa_H$ is defined as the average value of the normal curvatures $\kappa_H=(\kappa_1+\kappa_2)/2$. The *Gaussian curvature* $\kappa_G$ is defined as the product of the two principle curvatures $\kappa_G=\kappa_1\kappa_2$.

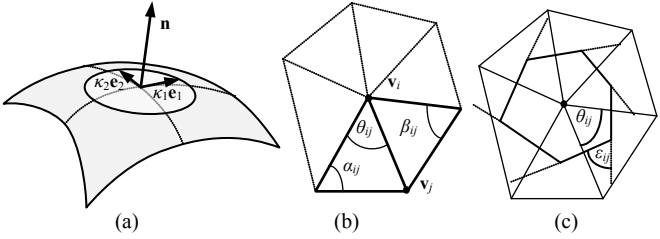

Figure 1. An illustration of (a) An infinitesimal neighborhood on a continuous surface patch; (b) 1-ring neighbors of point $v_i$ and angles opposite to an edge; (c) External angles of a Voronoi region.

For a given point $\mathbf{v}_i$ of discrete surface $M$ the mean curvature normal (Laplace-Beltrami operator) $\mathbf{K}(\mathbf{v}_i)=2\kappa_H(\mathbf{v}_i)\mathbf{n}(\mathbf{v}_i)$ gives both the mean curvature $2\kappa_H(\mathbf{v}_i)$ and unit normal $\mathbf{n}(\mathbf{v}_i)$ at the vertex $\mathbf{v}_i$. Expressions for the mean curvature normal and the Gaussian curvature of a discrete surface that depend only on a vertex position and angles of adjacent triangles respectively are given in [8]:

$$\mathbf{K}(\mathbf{v}_i) = \frac{1}{2\mathcal{A}_{Mixed}} \sum_{j\in N_1(i)} (cot\alpha_{ij}+cot\beta_{ij})(\mathbf{v}_j - \mathbf{v}_i)$$

$$\kappa_G(\mathbf{v}_i) = \frac{1}{\mathcal{A}_{Mixed}}\left(2\pi - \sum_{j=1}^{\#f}\theta_{ij}\right) \quad (1)$$

where $\#f$ is a number of adjacent triangular faces at the point $\mathbf{v}_i$, and $\theta_{ij}$ is the angle of $j$-th adjacent triangle at the point $\mathbf{v}_i$. $\mathcal{A}$ is the area of the first ring of triangles around the point $\mathbf{v}_i$. Minimizing a discretization error, Mayer, Desbrun, Schroder, and Barr [8] suggest a region $\mathcal{A}_{Mixed}$ as a combination of Voronoi region for non-obtuse triangle and the half (or quarter) of an obtuse triangle region (Fig. 2.).

$$\mathcal{A}_{Voronoi} = \frac{1}{8}\sum_{j\in N_1(i)}(cot\alpha_{ij}+cot\beta_{ij})\|v_i-v_j\|^2 \quad (2)$$

$$\mathcal{A}_{non-Vor.} = \begin{cases} \frac{1}{4}\sum_{j\in N_1(i)}(cot\theta_{ij}+cot\alpha_{ij})\|v_i-v_j\|^2 \sin^2\theta_{ij} & a) \\ \frac{1}{8}\sum_{j\in N_1(i)}(cot\theta_{ij}+cot\alpha_{ij})\|v_i-v_j\|^2 \sin^2\theta_{ij} & b) \end{cases} \quad (3)$$

The expressions for discrete principal curvatures at the point $\mathbf{v}_i$ are:

$$\kappa_1(\mathbf{v}_i) = \kappa_H(\mathbf{v}_i)+\sqrt{\Delta(\mathbf{v}_i)}, \quad \kappa_2(\mathbf{v}_i) = \kappa_H(\mathbf{v}_i)-\sqrt{\Delta(\mathbf{v}_i)} \quad (4)$$

where are

$$\Delta(\mathbf{v}_i) = \kappa_H^2(\mathbf{v}_i)-\kappa_G(\mathbf{v}_i), \quad \kappa_H(\mathbf{v}_i) = \frac{1}{2}\|\mathbf{K}(\mathbf{v}_i)\|. \quad (5)$$

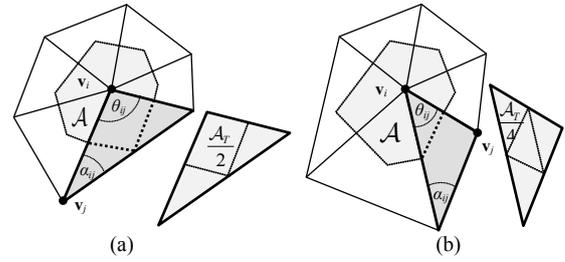

Figure 2. An example of mixed areas around a vertex use the midpoint of the edge opposite to the obtuse angle as the center: (a) $\mathcal{A}_{non-Voronoi} = \mathcal{A}_T/4$ and (b) $\mathcal{A}_{non-Voronoi} = \mathcal{A}_T/2$.

*2) Fitting Quadric Curvature Estimation*

Fitting Quadric Curvature Estimation is one of the most used methods for curvature features computation today. McIvor and Valkenburg give mathematical and theoretical background and comparison to previous similar methods in [9]. Later, Petitjean published survey [10] with comparison to the discrete differential geometry method, whereas Toolbox Graph (free Matlab software tools by Piere Alliez) is the practical application of this method; author refers to detailed algorithm in [11] and [12]. The idea of this method is that smooth surface geometry can be locally approximated with a quadratic polynomial surface. Method fits a quadric to point in a local neighborhood of each chosen point in a local coordinate frame. The frame is positioned at the chosen point and with the Z-coordinate axis, which is aligned along an estimated surface normal at that point. The curvature of the quadric at the chosen point is taken to be the estimation of curvature for the discrete surface.

For a simple quadric form $z' = ax'^2+bx'y'+cy'^2$ procedure is given [9] by:

- Estimation of surface normal **n** at the point **v** by one of two ways: simple or weighted averaging, or finding a least squares fitted plane to the point and its neighbors.
- Positioning of a local coordinate system $(x',y',z')$ at the point **v** and aligning axis $z'$ along the estimated normal. McIvor and Valkenburg suggest aligning of the $x'$ coordinate axis with a projection of the global $x$ axis onto the tangent plane defined by **n**. This results in *attitude matrix* rotation

$$\mathcal{R}' = [\mathbf{r}_1, \mathbf{r}_2, \mathbf{r}_3]^T \quad (6)$$

- from the global coordinate frame to the local coordinate frame. Then:

$$\mathbf{r}_3 = \mathbf{n} \quad \mathbf{r}_1 = \frac{(\mathcal{I}-\mathbf{n}\mathbf{n}^T)\mathbf{i}}{\|(\mathcal{I}-\mathbf{n}\mathbf{n}^T)\mathbf{i}\|} \quad \mathbf{r}_2 = \mathbf{r}_3 \times \mathbf{r}_1 \quad (7)$$

where $\mathcal{I}$ is the identity matrix, and **i** is the global $x$ axis $[1,0,0]^T$.

- Mapping the world data to the rotated principal frame with: $\mathbf{x}' = \mathcal{R}'(\mathbf{x}_w - \mathbf{v}_w)$, where index $_w$ denotes that coordinates are expressed in global (world) coordinate frame.
- Fitting the mapped data to the rotated principal quadric $z'= a'x'^2+b'x'y'+c'y'^2$, and solve the resulting system of linear equations, giving $a',b',c'$:

$$\begin{bmatrix} x_1^2 & x_1y_1 & y_1^2 \\ \vdots & \vdots & \vdots \\ x_n^2 & x_ny_n & y_n^2 \end{bmatrix} \begin{bmatrix} a' \\ b' \\ c' \end{bmatrix} = \begin{bmatrix} z_1 \\ \vdots \\ z_n \end{bmatrix} \quad (8)$$

- Computing both principal curvatures $\kappa_1$ and $\kappa_2$ using:

$$\kappa_1 = a' + c' + \sqrt{(a'-c')^2 + b'^2}$$
$$\kappa_2 = a' + c' - \sqrt{(a'-c')^2 + b'^2} \quad (9)$$

after that one can give expressions for Gaussian and mean curvature respectively:

$$\kappa_G = 4a'c' - b'^2, \quad \kappa_H = a' + c' \quad (10)$$

- Giving a new estimate of the surface normal at point **v**:

$$\mathbf{n} = \frac{[-d', -e', 1]^T}{1 + d'^2 + e'^2}. \quad (11)$$

if we use improvements suggested by McIvor and Valkenburg and fit the mapped data to extended quadric:

$$\hat{z} = a'\hat{x}^2 + b'\hat{x}\hat{y} + c'\hat{y}^2 + d'\hat{x} + e'\hat{y} \quad (12)$$

- Finally, we have estimation for Gaussian and mean curvature:

$$\kappa_G = \frac{4a'c' - b'^2}{(1+d'^2+e'^2)^2}, \kappa_H = \frac{a'+c'+a'e'^2+c'd'^2-b'd'e'}{\sqrt{(1+d'^2+e'^2)^3}} \quad (13)$$

where $a',b',c',d',e'$ are the parameters of the last quadric fitted.

### 3) Comparison of curvature estimation methods

A comparison of mentioned methods that have performed by McIvor and Valkenburg, and later Petitjean or Garimella and Swartz [13] gives enough information about accuracy. Our approach however does not require going into details.

More purposeful is experimental computation of characteristic curvature features using both methods, and checking stability of vertices with highest calculated values of the all curvature features. For our considerations, the method is better than other one if its most important vertices survive the deletion process of optimization.

Experimental and statistical results, presented in Appendix A, show different discrete values of characteristic curvature features for both methods (Table VI). Moreover, the difference of stable vertices appears, although only difference in their quantity is expected.

### B. Curvature estimation improvement

Roughly speaking, the basic difference between mentioned methods is a direction of computation of characteristic features. Thus, fitting quadric method first calculates both principal curvatures, and then derives Gaussian and mean curvature. The differential geometry method however begins calculation with the Gaussian curvature. In discrete case this feature is calculated as a sum of all adjacent triangle angles at vertex, and then divided by an amount of the selected surface area around the chosen vertex.

For our approach more important is the method with a direct impact on the calculation of Gaussian and mean curvature. Following our experimental results, we are going to improve the accuracy of curvature estimation for the method based on discrete differential geometry. We have assured that the local curvature computation using the Voronoi area did not give good experimental results; for obtuse angles, but also for any type of angles in strict local curvature estimation. Therefore, instead of Voronoi area we suggest using a *barycentric* area.

$$\mathcal{A}_{Barycenter} = \frac{1}{6} \sum_{j \in N_1(i)} \left( \cot\theta_{ij} + \cot\alpha_{ij} \right) \|x_i - x_j\|^2 \sin^2\theta_{ij} \quad (14)$$

Barycentric area at vertex $\mathbf{v}_i$ is calculated as a surface inside a closed contour formed by the midpoints of all edges belonging to the vertex $\mathbf{v}_i$ and the barycenters of its adjacent triangles. However, experimental results (See Appendix B) indicate a greater dependence of curvature estimation from the angle $\theta$ than the theoretical region convergence. Indeed, this argument is valid for measuring the stability of certain areas only, not for the global curvature evaluation.

### C. Optimization and Simplification

The second group of vertex features that we have used in consideration is a set of features that characterize salience of regions [14][15] and resistance to transformations. Since we know that rotation, translation and uniform scaling of 3D models do not affect the topological features of mesh vertices, optimization and simplification most destructive affect the object geometry. Therefore, it is necessary to understand principles of optimization and simplification for a reliable vertex importance determination. The commonly used optimization tools include the algorithms such as: Mesh Optimization [16], Progressive Meshes [17], Simplification Envelopes [18], Quadric Error Metrics [19], Memoryless Simplification [20], and Quadric-based simplification [21]. These processes drastically change the geometric structure, removing vertices and deforming triangular faces. Comparisons of simplification algorithms are given in [22] and [23].

The basic principle of 3D surfaces simplification, that is also the most relevant to this paper, is the vertex decimation process. Schroeder, Zarge, and Lorensen [24] give the details of this algorithm in a prominent paper "Decimation of triangle meshes". The algorithm begins by multiple passes of testing the local 3D mesh geometry over all its vertices. During the test each vertex of 3D mesh is a candidate for removal, but vertex and its surrounding triangular faces are deleted if tested vertex meets specified decimation criteria. Resulting hole in the mesh is patched by a local triangulation. The vertex removal process is repeated until a termination condition is met, and steps of the algorithm are:
- Characterization of the local vertex,
- The decimation criteria evaluation
- The resulting hole triangulation.

### 1) Local geometry characterization

The first step of decimation algorithm is a characterization of local geometry and topology for a chosen vertex. The outcome of this process determines whether the vertex is a potential candidate for deletion. Each vertex can be classified according to one of three possible classifications: simple, complex and boundary vertex. The simple vertex can be further classified as a smooth vertex, an interior edge or a corner vertex.

The *simple vertex* is surrounded by triangles where each edge containing the vertex is common to exactly two triangles. If the edge containing the vertex is shared with more than two triangles, or if vertex is common with at least one more than the cycle triangles, the *vertex is complex*. If an edge, which is branch of only one triangle, exists, then the vertex that belongs to this edge is the *boundary vertex*.

If the dihedral angle between two adjacent triangles is greater than a specified feature angle, then a feature edge exists. When a vertex is used by two feature edges, the vertex is an *interior edge vertex*. If three or more feature edges contain the same vertex, this common vertex is classified as the *corner vertex*. The illustration of characterization one may see in following figure [24]:

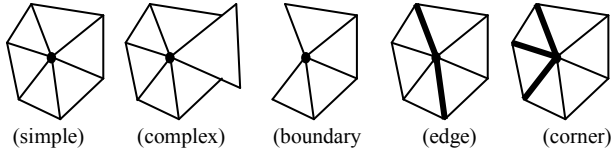

(simple)   (complex)   (boundary)   (edge)   (corner)

Figure 3. Classification of vertices according to their local geometric and topological features.

*2) The decimation criteria evaluation*

The evaluation step determines which the vertices and triangular faces can be deleted, or eventually replaced by another triangulation. The main criterion for the decimation of a simple vertex is its distance from the average plane formed by adjacent vertices. If this distance is less than a given one, then the vertex is deleted (Fig. 4. a). Boundary vertex and the interior edge decimation use the distance from the edge criteria. In this case, the algorithm determines the distance from the line, defined by the two vertices that create the boundary or interior edge. If the distance is less than a certain value, vertex can be deleted (Fig. 4. b).

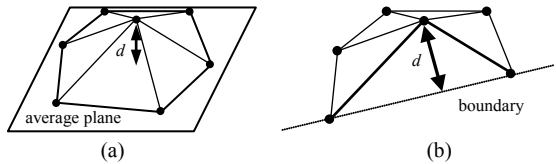

(a)          (b)

Figure 4. Criteria of decimation: a) vertex distance from average plane, b) vertex distance from boundary

Other simplification approaches are different from the specified vertex decimation approach in one, or even in all steps of algorithm. However, all of simplification algorithms minimize an energy function [25], and in this process some vertices, edges and faces are classified as risky primitives.

*3) Risky primitives and topological errors*

There is the list of risky geometric primitives:
- Isolated vertices, independent in space,
- vertices which belong to only one boundary edge,
- complex vertices and edges that join more than two faces,
- crossed edges without mutual vertex
- vertices with highest quantity of neighbors,
- vertices that are connected by collinear edges,
- boundary vertices, edges and faces,
- vertices, edges and faces in flat and smooth areas ($d\rightarrow0$).

This listed group of primitives can be divided in two sub-groups: topological errors (the first four) and regular risky primitives (the last four). Topological errors are treated by our assessment criteria with strict zero grade, while regular risky primitives require different treatment and their assessment criteria should be adjustable to user.

*D. Curvature sign*

Based on the calculated Gaussian curvature, both principal curvatures and relationship between them, one can define vertex features in all areas of a given mesh *M*. The point $v_i \in M$ is referred to as an *umbilics* if $\kappa_1(v_i)=\kappa_2(v_i)$. If $\kappa_1(v_i)=\kappa_2(v_i)=0$, $v_i$ is noted as a *planar* point. If $\kappa_G(v_i)=0$, but $v_i$ is not a planar point, then $v_i$ is a *parabolic* point. If $\kappa_G(v_i)>0$, $v_i$ is an *elliptic* point, and if $\kappa_G(v_i)<0$, $v_i$ is called a *hyperbolic* point. Taking into account signs of both the Gaussian and mean curvatures we may distinguish the following eight basic surface shapes.

TABLE I. EIGHT BASIC LOCALLY SURFACE TYPES IN RESPECT WITH THEIR GAUSSIAN AND MEAN CURVATURES

| $\kappa$ | Vertex features | Surface shape | |
|---|---|---|---|
| $\kappa_G=0$ $\kappa_H=0$ | planar | planar | |
| $\kappa_G=0$ $\kappa_H<0$ | parabolic | ridge convex | |
| $\kappa_G=0$ $\kappa_H>0$ | parabolic | valley concave | |
| $\kappa_G>0$ $\kappa_H<0$ | elliptic | convex | |
| $\kappa_G>0$ $\kappa_H>0$ | elliptic | concave | |
| $\kappa_G<0$ $\kappa_H=0$ | hyperbolic | saddle | |
| $\kappa_G<0$ $\kappa_H<0$ | hyperbolic | saddle predominantly ridge convex | |
| $\kappa_G<0$ $\kappa_H>0$ | hyperbolic | saddle predominantly valley concave | |

Shape of shown areas determines the condition of the vertex stability, whose measure is defined by a value of the curvature; more precisely, by characteristic features of curvature.

### III. ORDERED STATISTICS VERTEX EXTRACTION AND TRACING ALGORITHM (OSVETA) [28]

The essence of OSVETA is represented by three steps: defining and ranking assessment criteria, accurate curvature evaluating and computing its characteristic features, as well as tracing the importance of extracted vertices in relation to the mesh topology. Based on the previous theoretical consideration, the algorithm first extracts all of the important geometric features. Statistical results of experimental tests are however crucial for criteria ranking, and finalizing a list of assessment criteria. Roughly speaking, the most important question is: which highest grades are the most significant?

*A. Assessment criteria ranking*

The following table gives the answer to the question and shows extracted criteria, and appropriate rates:

TABLE II. OSVETA ASSESSMENT CRITERIA AND THEIR RATING

| $N^o$ | Criterion | Description | Rate |
|---|---|---|---|
| 1 | $\psi_{min} >= 0$ | Positive minimal dihedral angle | 1.0 |
| 2 | $\theta < 2\pi$ | Small Theta angle | 1.0 |
| 3 | $\kappa_{G1} > 0$ | Positive Gaussian curvature | 1.0 |
| 4 | $\psi_{max} >= 0$ | Positive maximal dihedral angle | 0.9 |
| 5 | $\theta > 2\pi$ | Big Theta angle | 0.8 |
| 6 | $\kappa_G < 0$ | Negative Gaussian curvature | 0.8 |
| 7 | $\kappa_{G1} < 0$ | Negative Gaussian curvature | 0.7 |
| 8 | $\kappa_G > 0$ | Positive Gaussian curvature | 0.4 |

$\psi_{max}$ and $\psi_{min}$ denote maximal and minimal dihedral angles at the vertex respectively. The angle $\theta$ is the sum of all $\theta_{ij}$ angles at the vertices (see Fig. 1. b). $\kappa_G$ and $\kappa_{GI}$ represent Gaussian curvatures estimated using respectively the both of method: discrete differential geometry method, and fitting quadrics method. The criteria ranking is performed using an automatic iterative computation of the survival vertices for the certain criteria selection. The results of tests are shown in Appendix A (See Table V).

### B. Irrelevant and Risky vertices elimination

The elimination of some *unimportant* vertices from the assessment, and their sorting is also a significant process. Irrelevant and risky vertices are isolated from further computations and an outcome of elimination is decreasing of irrelevant vertices assessment probability. In addition, it reduces chance that topological errors to be considered in calculations; as it is mentioned previously, this kind of vertices (See Section II.C.3) can get good grades for an eventual high value of curvature.

Hence, in this step the algorithm extracts the matrix of topological error vertices $E_M$, the matrix of boundary vertices $B_M$, and a vector of adjacent vertices: $N_A$ for each vertex $v_i$. Using these matrices as well as the sign and a threshold value[1] of $\kappa_G(v_i)$ and $\kappa_H(v_i)$ we first extract all *risky* vertices. The matrix of risky vertices is actually a union of matrices $E_M$, $B_M$[2] and a set of vertices which satisfy the condition: $\kappa_G(v_i) = 0 \wedge \kappa_H(v_i) = 0$.

In Table III we introduce new criteria with geometric features, obtained from conclusions of the consideration in Section II.C.: $C_{el}$ is a coefficient of maximal elongated face; $C_{DUG}$ is a length of the longest connected edge to the vertex; $C_{TUP}$ is a value of the maximal angle of adjacent triangles; $C_{VIS}$ is a distance of the vertex from the average plane. $\nabla \kappa_G$ and $\nabla \kappa_H$ are gradients of Gaussian and mean curvature respectively.

TABLE III. IRRELEVANT VERTICES ELIMINATION CRITERIA

| $N^0$ | Criterion |
|---|---|
| 1 | ($\kappa_G > 0$ & $\kappa_G < 7.0513e-13$) | ($\kappa_G < 0$ & $\kappa_G > -2.8223e-9$) |
| 2 | $\kappa_H > 2.3688$ | ($\kappa_H < 6.0769e-6$ & $\kappa_H > 0$) |
| 3 | $\kappa_H < -51929.7144$ | ($\kappa_H > -3.8486e-6$ & $\kappa_H < 0$) |
| 4 | $\kappa_{max} > 39.5904$ |
| 5 | $\kappa_{max} < -594.6627$ | ($\kappa_{max} < 0$ & $\kappa_{max} > -2.3077e-15$) |
| 6 | $\kappa_{min} > 0.40945$ | ($\kappa_{min} > 0$ & $\kappa_{min} < 0.00017504$) |
| 7 | $\kappa_{min} < -51843.002$ | ($\kappa_{min} < 0$ & $\kappa_{min} > -3.6061e-6$) |
| 8 | $\theta > 545.3552$ | $\theta < 213$ | $\theta == 360$ |
| 9 | $\kappa_{GI} > 0.83844$ | $\kappa_{GI} < -1.2989$ |
| 10 | $\kappa_{HI} > 1.2481$ | ($\kappa_{HI} >= 0$ & $\kappa_{HI} < 1.671e-5$) |
| 11 | $\kappa_{HI} < -2.3336$ | ($\kappa_{HI} < 0$ & $\kappa_{HI} > -1.5326e-8$) |
| 12 | $C_{el} > 1.7682$ | ($C_{el} > 0$ & $C_{el} < 8.4903e-7$) |
| 13 | $C_{el} < -1.3404$ | ($C_{el} < 0$ & $C_{el} > -6.9662e-5$) |
| 14 | $C_{DUG} > 6.1607$ | $C_{DUG} < 1.0111$ |
| 15 | $C_{TUP} > 179.9596$ | $C_{TUP} < 32.7365$ |
| 16 | $C_{VIS} > 0.92554$ | $C_{VIS} < 3.7982e-6$ |
| 17 | $\nabla \kappa_G > 1467169.264$ | $\nabla \kappa_G < 6.0536e-17$ |
| 18 | $\nabla \kappa_H > 2.8209$ | $\nabla \kappa_H < 5.6712e-5$ |
| 19 | $\psi_{max} >= 90$ | $\psi_{max} < 2$ |

---
[1] Threshold values of all elimination criteria, including Gaussian and mean curvature values, are adjustable. In the OSVETA software [28], we allow three setting options: safe, extended and aggressive.

[2] Boundary vertices are often important for shape creation, and algorithm leaves to an user the choice of their elimination from ordering.

### C. The flowchart of Algorithm process

The flowchart of all OSVETA algorithm processes is shown on Fig. 5.

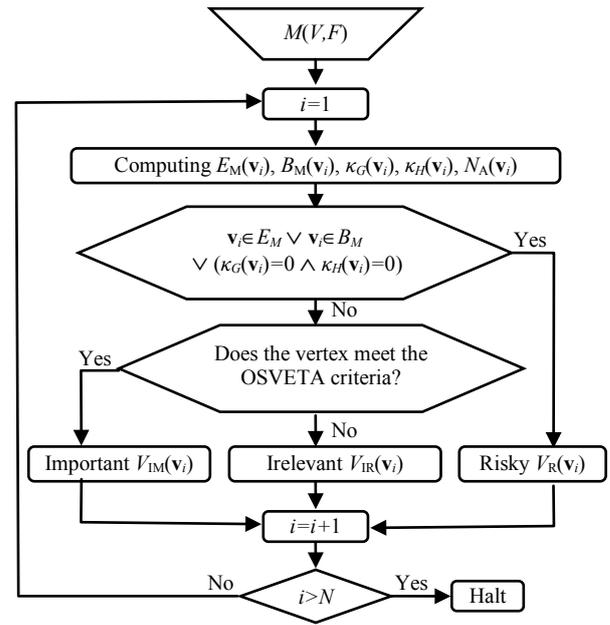

Figure 5. The flow chart of the OSVETA.

For a mesh of a given 3D surface $M(V,F)$ the result of OSVETA are two vectors: **s**, the vector of vertex stabilities arranged in a decreasing order, and **i**, the vector of corresponding indices. The mesh vertices are ordered with respect of decreasing stability form the vector $V_o$. $V_o=v_i$, and the length-$L$ vector **p** is obtained by taking the first $L$ elements from $V_o$. $N$ denotes the total number of vertices and also the total number of iterations for the shown process.

## IV. NUMERICAL RESULTS

In this section we present a practical usability of our software in computation of the vertex importance, i.e. vertex invariance to the optimization process. We have used four 3D mesh models (See Fig. 6.):
(A) Christ the Redeemer [27], (B) Myron of Eleutherae [27], (C) Naissa by Bata [26], (D) Venus de Milo [27].

For brevity, we refer to these objects as to A, B, C, and D. All four models are different in total number of vertices and faces, but also in geometric structure. More precisely, they are different in percentage of curved and flatten areas. Myron of Eleutherae and Naissa by Bata are complex mesh models; both containing closed elements as sub-objects. OSVETA algorithm works equally well with both homogeneous and complex meshes, as well as with open meshes with a boundary.

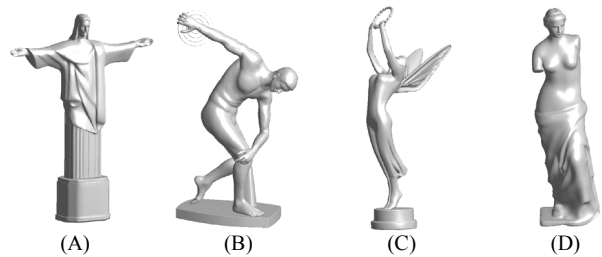

(A)      (B)      (C)      (D)

Figure 6. Original 3D meshes: (A) Christ the Redeemer, (B) Myron of Eleutherae, (C) Naissa by Bata [18] and (D) Venus de Milo.

In order to compare OSVETA with a method of random vertex selection, we have used 'Optimize' modifier from 3D Studio Max 2012 application [29]. We have performed seven different levels of optimization with the following face thresholds:

(A)(FT): 0,2,4,6,8,10,13,25  (B)(FT): 0,2,4,6,8,10,13,27
(C)(FT): 0,2,4,6,8,10,13,26  (D)(FT): 0,2,4,6,8,10,13,24

Higher maximal values are used for meshes with the larger total number of faces. The zero FT value does not affect the reduction of vertices and faces, and we consider it as an initial state of mesh geometry. The optimization with the maximum FT value destroys completely the geometric structure of the mesh, leaving only 5-10% of the total number of vertices (See Fig. 7.).

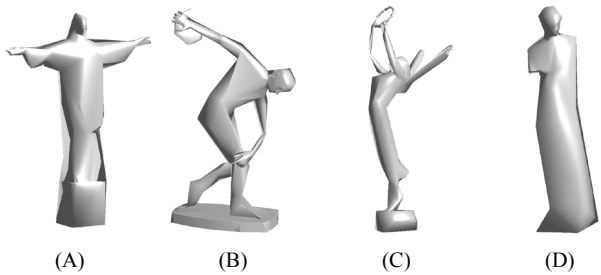

(A)      (B)      (C)      (D)

Figure 7. Destroyed geometries of 3D meshes after optimization with high level of FT.

A determination of face threshold limits, preserving the usability of 3D mesh, is illustrated in Fig. 8. in the example of the Venus de Milo mesh. As can be seen from Fig. 8. the optimization face threshold higher than 8 leads to significant perceptual and geometric degradation of mesh. Thus we set the OSVETA criteria for assessing vertices without considering FT optimization values over 10.

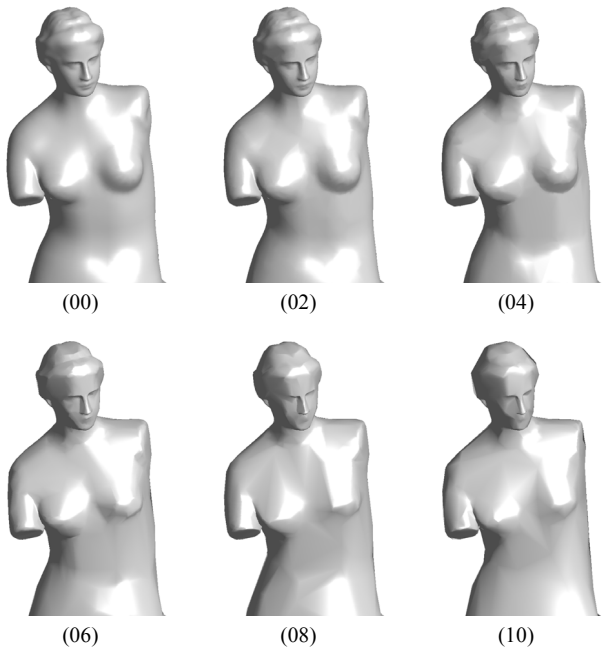

(00)      (02)      (04)

(06)      (08)      (10)

Figure 8. Perceptual degradation of the Venus de Milo 3D mesh as a function of the face threshold level. The subfigures correspond to the threshold levels 0,2,4,6,8,10.

Experimental tests of stability with 1000 vertices selected by our algorithm showed the superiority of our approach compared to the randomly allocated group of 1000 vertices. The ratio of removed vertices from mentioned mesh models, denoted by (A), (B), (C) and (D) numerically is shown in the Table IV. Taking into account that different vertices are removed due to different levels of optimization, the table shows the accurate matching undeleted vertices of our algorithm.

TABLE IV. RELATION OF NUMBER OF DELETED VERTICES BY OPTIMIZATION, USING RANDOM AND OSVETA EXTRACTION ALGORITHMS.

| 3D Model | FT=0 | FT=2 | FT=4 | FT=6 | FT=8 | FT=10 |
|---|---|---|---|---|---|---|
| A# Vertices | 27802 | 16627 | 9789 | 5991 | 3743 | 2432 |
| Random | 0 | 502 | 718 | 838 | 907 | 948 |
| OSVETA | 0 | 8 | 51 | 170 | 333 | 531 |
| B# Vertices | 100681 | 49931 | 32353 | 21498 | 15078 | 11620 |
| Random | 0 | 523 | 686 | 794 | 848 | 877 |
| OSVETA | 0 | 6 | 8 | 19 | 51 | 94 |
| C# Vertices | 33465 | 25334 | 17472 | 12146 | 8588 | 5870 |
| Random | 0 | 276 | 495 | 654 | 760 | 845 |
| OSVETA | 0 | 3 | 21 | 44 | 77 | 156 |
| D# Vertices | 17350 | 12209 | 6953 | 3926 | 2315 | 1448 |
| Random | 0 | 332 | 622 | 781 | 872 | 920 |
| OSVETA | 0 | 1 | 30 | 147 | 332 | 522 |

The one of most important features of our OSVETA extraction is stability of vertices grouped at the beginning of the selection. This fact is shown in the figure below used Naissa by Bata 3D mesh model, optimized using the limit level of face threshold (FT=10). Curve on the figure represents values of the Gaussian curvature at vertices selected using our algorithm (see Fig. 9.).

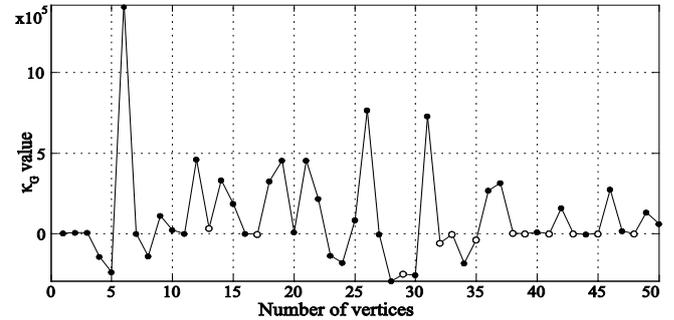

Figure 9. Gaussian curvature values at the first 50 extracted vertices by OSVETA: stabile vertices (filed circles) and deleted vertices by optimization (outlined circles).

We observed another important characteristic from OSVETA computations: probability of two consecutive vertices deletion is very low. For example, for all 3D mesh models, using all optimization levels, in 1000 extracted vertices there is no consecutive deleted vertices.

The presented vertex extraction approach based on OSVETA algorithm practically does not have restriction in use. A lack of such extraction is a dependence of curvature estimation accuracy, but we partially solve this problem using both evaluation methods as assessment criteria ($\kappa_G$ and $\kappa_{G1}$).

Although the computations of feature angles ($\theta$ and $\psi$) are used in Gaussian and mean curvature estimation, experimental results of feature angles criteria indicate a fairly more efficiency (related on the extracted vertices stability) than Gaussian and mean curvature criteria (See Table V in Appendix A).

## V. Conclusions and Future Work

We have elaborated a novel method of extraction of important vertices from a 3D mesh. In several steps of the algorithm, the geometric importance of mesh vertices are ranked by the criteria based on the curvature and other noted important features. Thus, the two crucial goals are realized: first, the vector of irrelevant and risky vertices is separated from the mesh; second, the vector of the ordered vertex stability, with corresponding indices is extracted.

Ordered Statistics Vertex Extraction and Tracing Algorithm have reached good efficiency showing invariance of extracted vertices to deletion by optimization. However, we could see that a success of algorithm is determined by results of two research directions: curvature estimation and simplification of the process characteristics and effects. Thus, OSVETA future improvements will involve more precise mesh geometry estimation and better curvature and topological feature estimation. We believe that these enhancements would result in more accurate identification of stable vertices and consequently significant reduction of deletion probability.

## Appendix A

In this part of the paper, we present comparison of numerical results, which we have obtained from an experimental test. For the computation test, we have used four mesh models (See Section IV).

The used models contain a different total number of vertices and faces. Geometric structure of each objects is also unique, i.e. the ratio of curved and flat areas is distinct for each model. In addition, Myron of Eleutherae and Naissa by Bata are complex mesh models, i.e. the both contain closed elements as sub-objects.

TABLE V. Criteria efficiency for given number of selected vertices.

| $N^0$ | Criterion | Number of selected vertices ||||||||| 
|---|---|---|---|---|---|---|---|---|---|---|
| | | 1 | 2 | 3 | 4 | 5 | 10 | 50 | 100 | 500 | 1000 |
| 1 | $\kappa_G > 0$ | 1 | 2 | 2 | 3 | 4 | 7 | 19 | 27 | 86 | 148 |
| 2 | $\kappa_G < 0$ | 0 | 1 | 2 | 3 | 4 | 7 | 21 | 31 | 86 | 128 |
| 3 | $\kappa_H > 0$ | 1 | 2 | 2 | 2 | 3 | 5 | 13 | 23 | 37 | 37 |
| 4 | $\kappa_H < 0$ | 1 | 1 | 1 | 1 | 1 | 2 | 4 | 9 | 34 | 69 |
| 5 | $\kappa_{max} > 0$ | 1 | 2 | 2 | 2 | 2 | 4 | 17 | 25 | 86 | 129 |
| 6 | $\kappa_{min} < 0$ | 0 | 0 | 0 | 1 | 1 | 1 | 4 | 9 | 42 | 80 |
| 7 | $\theta > 360$ | 0 | 1 | 1 | 2 | 2 | 7 | 31 | 54 | 138 | 192 |
| 8 | $\theta < 360$ | 1 | 1 | 2 | 3 | 4 | 7 | 31 | 47 | 130 | 185 |
| 9 | $\kappa_{G1} > 0$ | 1 | 2 | 3 | 3 | 3 | 5 | 21 | 38 | 99 | 131 |
| 10 | $\kappa_{G1} < 0$ | 0 | 0 | 0 | 0 | 1 | 3 | 17 | 31 | 82 | 125 |
| 11 | $\kappa_{H1} > 0$ | 0 | 1 | 1 | 1 | 1 | 2 | 11 | 18 | 79 | 116 |
| 12 | $\kappa_{H1} < 0$ | 0 | 1 | 1 | 2 | 2 | 2 | 12 | 20 | 68 | 99 |
| 13 | $C_{el} > 0$ | 0 | 0 | 0 | 0 | 1 | 1 | 6 | 10 | 41 | 70 |
| 14 | $C_{el} <= 0$ | 0 | 1 | 1 | 1 | 1 | 1 | 2 | 3 | 24 | 34 |
| 15 | $C_{VIS} > 0$ | 0 | 0 | 0 | 1 | 1 | 4 | 8 | 18 | 52 | 87 |
| 16 | $C_{TUP} > 0$ | 0 | 0 | 0 | 0 | 0 | 1 | 4 | 9 | 39 | 71 |
| 17 | $C_{DUG} > 0$ | 0 | 1 | 1 | 1 | 1 | 1 | 1 | 3 | 18 | 34 |
| 18 | $\nabla \kappa_H >$ mean | 0 | 1 | 1 | 1 | 1 | 1 | 9 | 14 | 45 | 86 |
| 19 | $\nabla \kappa_H <$ mean | 0 | 0 | 0 | 0 | 0 | 0 | 0 | 0 | 7 | 19 |
| 20 | $\nabla \kappa_G >$ mean | 1 | 2 | 2 | 3 | 3 | 5 | 13 | 27 | 84 | 151 |
| 21 | $\nabla \kappa_G <$ mean | 0 | 0 | 0 | 0 | 0 | 0 | 0 | 0 | 0 | 4 |
| 22 | $\psi_{max} >= 0$ | 0 | 0 | 1 | 1 | 1 | 2 | 11 | 18 | 112 | 204 |
| 23 | $\psi_{min} >= 0$ | 1 | 2 | 3 | 4 | 5 | 6 | 22 | 37 | 86 | 134 |

First, we test an efficiency of all criteria, i.e. we select and sort vertices of source mesh by 23 criteria and find for each criterion a number of selected vertices, matched to vertices of optimized object. The test is repeated for a different number of selected vertices. Previous table (Table V) shows results of the test for Venus de Milo 3D mesh object (FT=13).

In the next experimental test, we compare selections of all noted criteria. Each of 23 criteria selects ten vertices and Fig. 10. represents the histogram of frequent selected vertices.

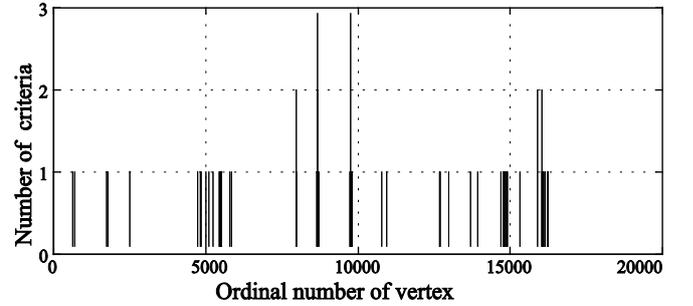

Figure 10. Frequency of the vertex selection (Venus de Milo).

As we can see, only two same vertices are selected by 3 criteria, three same vertices are selected by two criteria, and rests of vertices are selected by unique criterion.

The Histogram also indicates that the most of criteria selects unique vertices. Moreover, some similar criteria, e.g. Gaussian curvatures, estimated by two mentioned methods, select the complete different first 10 vertices. Table VI shows vertices selected by $\kappa_G > 0$ and $\kappa_{G1} > 0$ criteria for the maximal level of optimization (FT = 13).

TABLE VI. Differences of selected vertices by similar criteria.

| $N^0$ | Criterion | Selected vertices |||||||
|---|---|---|---|---|---|---|---|---|
| 1 | $\kappa_G > 0$ | 8755 | 8769 | 8773 | 9386 | 9439 | 10374 | 10376 |
| 9 | $\kappa_{G1} > 0$ | 5827 | 5829 | 6174 | 6833 | 15925 | | |

Indeed, geometric position of selected vertices in 3D space is the most important information. Thus, 3D plot on Fig. 11. shows both sets of selected vertices in their different topological regions. Perceptual and topological importance of regions to whom belong selected vertices is also clearly shown of next figure. We note that there is no identically selected vertex by both of criterion, i.e. vertices selected by each criterion are at different topological regions.

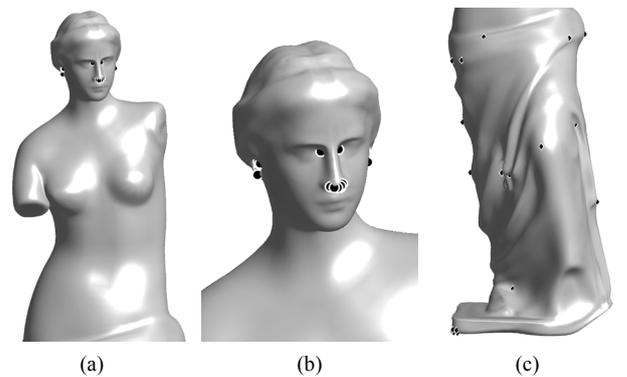

(a)     (b)     (c)

Figure 11. 3D position of first 50 vertices selected by $\kappa_G$ and $\kappa_{G1}$ criteria. (a) $\kappa_G > 0$ - circles, (b) zoom of (a), (c), $\kappa_{G1} > 0$ - diamonds.

APPENDIX B

We compute for Section II.B extraction efficiency of $\kappa_G > 0$ and $\kappa_G < 0$ criteria, using Voronoi and Barycentric regions in curvature estimation. For a better comparison with previous results Venus de Milo is used in extraction. Numbers of extracted vertices using both compared regions are given in following table:

TABLE VII. CRITERIA EFFICIENCY FOR GIVEN NUMBER OF SELECTED VERTICES.

| $N^0$ | Criterion | Number of selected vertices | | | | | | | | |
|---|---|---|---|---|---|---|---|---|---|---|
| | | 1 | 2 | 3 | 4 | 5 | 10 | 50 | 100 | 500 | 1000 |
| Curvature estimation using barycentric area in calculation | | | | | | | | | | | |
| 1 | $\kappa_G > 0$ | 1 | 2 | 2 | 3 | 4 | 7 | 19 | 27 | 86 | 148 |
| 2 | $\kappa_G < 0$ | 0 | 1 | 2 | 3 | 4 | 7 | 21 | 31 | 86 | 128 |
| Curvature estimation using Voronoi area in calculation | | | | | | | | | | | |
| 1 | $\kappa_G > 0$ | 1 | 2 | 2 | 3 | 4 | 6 | 18 | 25 | 84 | 140 |
| 2 | $\kappa_G < 0$ | 1 | 2 | 3 | 3 | 4 | 4 | 12 | 20 | 80 | 123 |

Significant differences in criteria efficiency using barycentric and Voronoi area in curvature estimations are observable only with large numbers of selected vertices. This fact is sufficient argument for use barycentric area in Gaussian and mean curvature calculation, because use of a small number of selected vertices are very rare.

In addition we may conclude that the use of barycentric areas in curvature calculations gives better results in the hyperbolic i.e. saddle shapes of mesh surfaces ($\kappa_G < 0$). In elliptic and parabolic regions ($\kappa_G > 0$ and $\kappa_G = 0$ respectively) curvature calculations, using both types of areas, ensure results with similar accuracy.

REFERENCES

[1] F. Attneave, "Some informational aspects of visual perception", *Psychological review*, vol. 61, no 3, pp. 183 – 193, 1954. [Online], Available: http://citeseerx.ist.psu.edu/viewdoc/summary?doi=10.1.1.212.633

[2] F. Attneave and M. D. Arnoult, "The quantitative study of shape and pattern perception," *Psychological bulletin*, vol. 53, no 6, pp. 452 - 471, 1956, [Online], Available: http://citeseerx.ist.psu.edu/viewdoc/summary?doi=10.1.1.211.6482

[3] P. Dombrowski, *150 years after Gauss' "Disquisitiones generales circa superficies curvas": with the original text of Gauss*, no 62, Société mathématique de France, 1979.

[4] A. D. Aleksandrov and V. A. Zalgaller, *Intrinsic geometry of surfaces*, vol. 15, of Translations of Mathematical Monographs, American Mathematical Society, 1967.

[5] J. J. Koenderink. Solid Shape, The MIT Press, Cambridge, Massachusetts, May 1990.

[6] J. P. Thirion, "The extremal mesh and understanding of 3D surfaces," *International Jurnal of Computer Vision*, vol. 19, no 2, pp. 115 - 128, August 1996. [Online], Available: http://rd.springer.com/article/10.1007/BF00055800

[7] M. Spivak. A comprehensive introduction to differential geometry, vol. 1, ED-3, Publish or Perish, January 1999.

[8] M. Meyer, M. Desbrun, P. Schröder, and A.H. Barr, "Discrete differential-geometry operators for triangulated 2-manifolds," *Visualization and Mathematics III*, ED-12, pp. 35 - 57, Springer, August 2003. [Online], Available: http://citeseerx.ist.psu.edu/viewdoc/summary?doi=10.1.1.24.3427

[9] A. M. McIvor and R. J. Valkenburg, "A comparison of local surface geometry estimation methods," *Machine Vision and Applications*, vol. 10, no 1, pp. 17 - 26, May 1997. [Online], Available: http://citeseerx.ist.psu.edu/viewdoc/summary?doi=10.1.1.45.2842

[10] S. Petitjean, "A Survey of methods for recovering quadrics in triangle meshes," *ACM Computing Surveys*, vol. 34, no 2, pp. 211 - 262, June 2002. [Online], Available: http://dl.acm.org/citation.cfm?id=508354

[11] D. Cohen-Steiner and J. M. Morvan. "Restricted Delaunay triangulations and normal cycle.", in *Proc. 19th Ann. Symp. Computational geometry - SCG '03*, San Diego, June 2003, pp. 312 - 321. [Online], Available: http://dl.acm.org/citation.cfm?id=777839

[12] P. Alliez, D. Cohen-Steiner, O. Devillers, B. Levy, and M. Desbrun, "Anisotropic polygonal remeshing," *ACM Transactions on Graphics*, vol. 22, no 3, pp. 485 - 493, July 2003. [Online], Available: http://dl.acm.org/citation.cfm?id=882296

[13] R. V. Garimella and B. K. Swartz, "Curvature Estimation for Unstructured Triangulations of Surfaces," Technical Report LA-UR-03-8240, Los Alamos National Laboratory, Los Alamos, New Mexico, USA, 2003.

[14] K. Watanabe and A. G. Belyaev, "Detection of Salient Curvature Features on Polygonal Surfaces," in *Proc. Computer Graphics Forum - EUROGRAPHICS 2001*, Manchester, September 2001, vol. 20, no 3, pp. 385 – 392. [Online], Available: http://onlinelibrary.wiley.com/doi/10.1111/1467-8659.00531/pdf

[15] C. H. Lee, A. Varshney, and D. W. Jacobs, "Mesh saliency," *ACM Transactions on Graphics*, vol. 24, no 3, pp. 659–666, July 2005. [Online], Available: http://citeseerx.ist.psu.edu/viewdoc/summary?doi=10.1.1.225.8964

[16] H. Hoppe, T. DeRose, T. Duchamp, J. McDonald, and W. Stuetzle, "Mesh optimization," in *Proc. 20th Annu. Conf. Computer graphics and interactive techniques - SIGGRAPH '93*, Anaheim, August 1993, pp. 19-26. [Online], Available: http://citeseerx.ist.psu.edu/viewdoc/summary?doi=10.1.1.46.4677

[17] H. Hoppe, "Progressive meshes," in *Proc. 23rd Annu. Conf. Computer graphics and interactive techniques - SIGGRAPH '96*, New Orleans, August 1996, pp. 99-108. [Online], Available: http://citeseerx.ist.psu.edu/viewdoc/summary?doi=10.1.1.153.3865

[18] J. Cohen, A. Varshney, D. Manocha, G. Turk, H. Weber, P. Agarwal, F. Brooks, and W. Wright, "Simplification envelopes," in *Proc. 23rd Annu. Conf. Computer graphics and interactive techniques - SIGGRAPH '96*, New Orleans, August 1996, pp. 119 - 128. [Online], Available: http://dl.acm.org/citation.cfm?id=237220

[19] M. Garland and P.S. Heckbert, "Surface simplification using Quadric Error Metrics," in *Proc. 24th Annu. Conf. Computer graphics and interactive techniques - SIGGRAPH '97*, Los Angeles, August 1997, pp. 209 - 216. [Online], Available: http://citeseerx.ist.psu.edu/viewdoc/summary?doi=10.1.1.41.7644

[20] P. Lindstrom and G. Turk, "Fast and memory efficient polygonal simplification," in *Proc. Visualization '98*, Research Triangle Park, October 1998, IEEE Computer Soc. Press, pages 279 – 286. [Online], Available: http://ieeexplore.ieee.org/xpl/articleDetails.jsp?arnumber=745314

[21] M. Garland and Y. Zhou, "Quadric-based simplification in any dimension," *ACM Transactions on Graphics*, vol. 24, no 2, pp. 209 - 239, April 2005. [Online], Available: http://citeseerx.ist.psu.edu/viewdoc/summary?doi=10.1.1.78.1270

[22] P. Heckbert and M. Garland, "Survey of polygonal surface simplification algorithms," *Multiresolution Surface Modeling Course - SIGGRAPH '97*, Los Angeles, August 1997. [Online], Available: http://citeseerx.ist.psu.edu/viewdoc/summary?doi=10.1.1.17.9821

[23] P. Cignonia, C. Montania, and R. Scopignob, "A comparison of mesh simplification algorithms," *Computers & Graphics*, vol. 22, no 1, pp. 37 – 54, February 1998. [Online], Available: http://citeseerx.ist.psu.edu/viewdoc/summary?doi=10.1.1.43.8837

[24] W. J. Schroeder, J. A. Zarge, and W. E. Lorensen, "Decimation of triangle meshes," in *Proc. 19th Annu. Conf. Computer graphics and interactive techniques - SIGGRAPH '92*, Chicago, July 1992, pp. 65 – 70. [Online], Available: http://citeseerx.ist.psu.edu/viewdoc/summary?doi=10.1.1.151.2200

[25] D. Zorin, "Curvature-based energy for simulation and variational modeling," in *Proc. International IEEE Conf. on Shape Modeling and Applications - SMI '05*, Cambridge, June 2005, pp. 196 - 204. [Online], Available: http://dl.acm.org/citation.cfm?id=1097876.1098471&coll=DL&dl=GUIDE&CFID=128134297&CFTOKEN=40865769

[26] Serbian Film Center, Naissa Trophy [Online], Available: http://silicon-studio.com/Naissa_by_Bata.zip

[27] Autodesk free 3D models, Autodesk 123C, [Online], Available: http://www.123dapp.com/123C-3D-Model/Venus-de-Milo-Statue/595750
http://www.123dapp.com/123C-3D-Model/Myron-of-Eleutherae-Statue/595747
http://www.123dapp.com/123C-3D-Model/Christ-the-Redeemer-Statue/595949

[28] B. Vasic, Ordered Statistic Vertex Extraction and Tracing Algorithm (OSVETA) - MatLab Software, [Online], Available: http://silicon-studio.com/OSVETA.zip

[29] M. Kauffman, "Optimizing Your Autodesk 3ds Max Design Models for Project Newport", *Autodesk University 2009*, [Online], Available: http://au.autodesk.com/?nd=material&session_material_id=6296,pp:6